\documentclass[a4paper,11pt,oneside]{article}

\usepackage[english]{babel}
\usepackage[T1]{fontenc}
\usepackage[utf8]{inputenc}

\usepackage[pdftex,unicode]{hyperref}
\hypersetup{pdftitle=Disaggregating Input--Output Tables by the Multidimensional RAS Method: A Case Study of the Czech Republic}
\hypersetup{pdfauthor=Vladimír Holý and Karel Šafr}

\usepackage{color}
\definecolor{mycolor}{rgb}{0.612,0.098,0.129}
\hypersetup{colorlinks=true, linkcolor=mycolor, anchorcolor=mycolor, citecolor=mycolor, filecolor=mycolor, urlcolor=mycolor}

\usepackage[margin=60pt]{geometry}
\usepackage{bm}
\usepackage{amsmath}
\usepackage{amssymb}
\usepackage{amsthm}
\newtheorem{lemma}{Lemma}
\newtheorem{theorem}{Theorem}

\usepackage[group-minimum-digits=3]{siunitx}

\usepackage{graphicx}
\usepackage{float}
\usepackage{hhline}
\usepackage{multirow}
\usepackage{booktabs}

\usepackage[authoryear]{natbib}

\newcommand{\bmni}[1]{\bm{\mathrm{#1}}}

\begin{document}

\begin{center}
{\Large \bfseries Disaggregating Input--Output Tables \\ by the Multidimensional RAS Method: \\ A Case Study of the Czech Republic}
\end{center}

\begin{center}
{\bfseries Vladimír Holý}\footnote{Corresponding author.} \\
Prague University of Economics and Business \\
Winston Churchill Square 4, 130 67 Prague 3, Czech Republic \\
\href{mailto:vladimir.holy@vse.cz}{vladimir.holy@vse.cz}
\end{center}

\begin{center}
{\bfseries Karel Šafr} \\
Prague University of Economics and Business \\
Winston Churchill Square 4, 130 67 Prague 3, Czech Republic \\
\href{mailto:karel.safr@vse.cz}{karel.safr@vse.cz}
\end{center}

\begin{center}
{\itshape \today}
\end{center}

\noindent
\textbf{Abstract:}
The RAS method is an iterative procedure that bi-proportionally scales an input--output table to be consistent with given row and column sums.
It can be used to disaggregate an annual national table to more detailed tables, such as regional, quarterly, and domestic/imported tables.
However, the regular two-dimensional RAS method does not ensure the consistency of the disaggregated tables with the original table.
For this problem, we use the multidimensional RAS method, which besides input and output totals, also ensures regional, quarterly, and domestic/imported totals.
Our analysis of Czech industries shows that the multidimensional RAS method increases the accuracy of table estimation as well as the accuracy of the Leontief inverse, the quarterly value added, and (to some degree) the regional Isard's model.
We also rigorously demonstrate its relation to the cross-entropy model.
\\

\\

\section{Introduction}
\label{sec:intro}

In economics, the relations between different branches of national economy can be described using input--output analysis. The key to this analysis is the input--output table, which shows how an output of one industry is an input to another industry.

It happens quite often in empirical work that an input--output table needs to be updated (or balanced) to be consistent with given row and column sums. A simple yet powerful method to obtain such an updated (balanced) input-output table is the RAS method. We describe the RAS method and its extensions in Section \ref{sec:litRas}. Notably, \cite{Tilanus1976a} and \cite{Cole1992} discuss the extension of the RAS method to multiple dimensions.

Our focus is the disaggregation of input--output tables.
Input--output tables can be disaggregated in many dimensions, such as space (i.e., a national table divided into regional tables), time (i.e., an annual table divided into intra-year tables), and division between domestic and imported use.
These disaggregated tables bring more insight to input--output analysis and have been the focus of many papers. We briefly review the literature about input--output disaggregation in Section \ref{sec:litDisagg}.

Following \cite{Tilanus1976a} and \cite{Cole1992}, we use the multidimensional RAS method for the compilation of disaggregated input--output tables.
Instead of two-dimensional input--output tables, we work with multidimensional tables.
We establish this setting in Section \ref{sec:methTables}.
The additional dimensions can represent space, time, or domestic/imported decomposition.
In this setting, the multidimensional RAS method simultaneously estimates all the input--output tables given by the additional dimensions, ensuring the totals in all dimensions to match the desired values.
For example, the multidimensional RAS method ensures that the sum of quarterly tables is equal to the annual table.
This is the main advantage of the multidimensional approach over the two-dimensional RAS method, which ensures only row and column sums to match the desired values even when used for the estimation of disaggregated tables.
Additionally, as the multidimensional RAS uses more information than its two-dimensional counterpart, it can be expected to be more accurate (see e.g., \citealp{DeMesnard2006}).
We formally describe the algorithm of the multidimensional RAS method in Section \ref{sec:methRas}.

The first contribution of our paper is to relate the multidimensional RAS method to the cross-entropy model.
We present the multidimensional cross-entropy model in Section \ref{sec:methEntropy} and rigorously show in Section \ref{sec:methRelation} that its solution is the same as that of the multidimensional RAS method.
Our proof is a multidimensional extension of the two-dimensional version of \cite{Lemelin2013}.

The second contribution lies in applying the multidimensional RAS method to the input--output tables of the Czech economy, showing the scope of its use, and demonstrating its benefits on real data.
In Sections \ref{sec:appSingle} and \ref{sec:appInter}, we apply the method to the national input--output table divided into regional tables.
Besides the comparison with the benchmark regional tables, we explore the utility of the method in the contexts of the Leontief inverse and the Isard's model.
In Section \ref{sec:appQuarterly}, we deal with the annual input--output tables divided into quarterly tables.
The multidimensional RAS method increases the accuracy of quarterly estimates of the value added for the years studied.
Lastly, in Section \ref{sec:appDomimp}, we show that the third dimension can also capture disaggregation of the input--output table based on domestic and imported use.
All these applications show that the error caused by omitting the third dimension is quite significant and the multidimensional RAS method also increases the accuracy of the estimated tables.

Note that the problems we treat in the multidimensional setting can also be treated by other methods.
The KRAS method of \cite{Lenzen2009} is a versatile tool which can accommodate conflicting information, negative elements, and arbitrary constraints, including those we label as multidimensional.
It follows that the multidimensional RAS method can be used only for a subset of the problems solvable by the KRAS method.
However, in our opinion, the multidimensional formulation of such specific problems is more natural and elegant.
The multidimensional RAS approach is also used by \cite{Temursho2021} and \cite{Valderas-Jaramillo2021} in the more general GRAS setting allowing for negative elements\footnote{We should note that the first version of our paper \cite{Holy2017l} predates \cite{Temursho2021} and \cite{Valderas-Jaramillo2021}.}.
Note that our approach is a special case of the multidimensional GRAS setting.
Nevertheless, our two contributions remain relevant as the theoretical properties of the more general methods have not yet been much explored and, in addition, our empirical findings are also applicable to the more general approaches.

\section{Literature Review}
\label{sec:lit}

\subsection{The RAS Method}
\label{sec:litRas}

The earliest use of the RAS algorithm for the estimation of input--output tables includes \cite{Stone1962b}, \cite{Stone1962}, and \cite{Bacharach1970}.
The purpose of the RAS method is to modify the elements of an input--output table in a way that is consistent with some predefined row and column totals.
The iterative algorithm starts with the original fully-known table.
In the first iteration it multiplies the rows of the original table so that their totals will be the same as the desired row totals and then it multiplies its columns so that their totals will be the same as the desired column totals.
However, the multiplication of the columns leads to a violation of the row totals, and vice versa, so several iterations must be performed.
For the algorithm to converge, it is required that the row and column totals be consistent, i.e., the sum of the row totals be equal to the sum of the column totals.
The matrix must also have a ``suitable'' structure in terms of zero elements (see, e.g., \citealp{Macgill1977} for a rigorous treatment).
When all elements of the matrix are strictly positive, the convergence is ensured (provided the totals were consistent).
 
The theoretical properties of the RAS method were reviewed by \cite{Macgill1977}.
Proofs of convergence under various assumptions were presented for example by \cite{Bacharach1965}, \cite{Sinkhorn1967}, and \cite{Evans1970} using calculus and linear algebra, by \cite{Fienberg1970} using differential geometry, and by \cite{Ireland1968}, \cite{Csiszar1975}, and \cite{Ruschendorf1995} using the entropy approach.
\cite{Pukelsheim2014} dealt with the necessary and sufficient conditions for convergence.
The complexity and effective implementation of the algorithm was studied by, e.g., \cite{Jirousek1995} and \cite{Kalantari2008}.

The RAS method is known under many other names in different scientific fields.
Originally, \cite{Deming1940} introduced the iterative proportional fitting procedure for the estimation of the values of the cells of contingency tables,
stating that the algorithm minimizes Pearson's $\chi^2$ statistic, but, as shown by \cite{Stephan1942}, this is not correct.
Another application is the trip distribution problem, in which the RAS algorithm is known as the Furness method \citep{Evans1970}.
It is also known as biproportional fitting, iterative proportional scaling, or matrix raking in statistics, while computer scientists prefer the term matrix scaling.
The RAS method is closely related to the cross-entropy as studied by \cite{Mcdougall1999} and \cite{Lemelin2013}.
Having many different names for basically the same method has caused confusion in the literature, as many authors have been unaware of the work in different fields.
This was noted by, for example, \cite{Johnston1993}, who discussed iterative proportional fitting procedure and cross-entropy in the geographical context with an illustration by British voting patterns.

The RAS method has received several extensions over the years.
The modified RAS (MRAS) method of \cite{Lecomber1975} deals with  uncertainty in the preliminary estimates.
Case studies were presented by \cite{Allen1974} and \cite{Allen1975}.
\cite{Cole1992} extended the RAS method to multiple dimensions.
\cite{Gilchrist1999} proposed a three-stage extension of the RAS algorithm, called the three-stage RAS (TRAS).
It allows adding restrictions to arbitrary subsets of matrix elements in addition to the row and column totals.
The generalized RAS (GRAS) algorithm proposed by \cite{Junius2003} generalizes the RAS method by allowing matrices with some negative elements.
A corrected exposition of the GRAS method was presented later by \cite{Lenzen2007}.
\cite{Temurshoev2013} improves this method by allowing some rows or columns to consist only of non-positive elements.
\cite{Lemelin2009} suggests a GRAS variant based on minimum information loss.
The \textit{konfliktfreies} RAS (KRAS) method proposed by \cite{Lenzen2009} incorporated features of TRAS, MRAS, and GRAS, and further generalized the RAS algorithm to the case of conflicting data.
Furthermore, \cite{Lenzen2014} present a non-sign-preserving modification of GRAS and KRAS methods.
Recently, \cite{Temursho2021} and \cite{Valderas-Jaramillo2021} extended the GRAS method to multiple dimensions.
\cite{Mesnard2021} discussed alternative generalizations of the two-dimensional RAS problem to multiple dimensions.
A more detailed overview of extensions of RAS as well as the history of the developement of the RAS method can be found in \cite{Lahr2004}.
Recent applications of the RAS method include \cite{Alvarez-Martinez2018}, \cite{Cai2019}, \cite{Okuyama2019}, and \cite{FournierGabela2020}.

While the RAS method seems to be the most popular approach for adjusting a matrix with given row and sums constraints, there are some alternatives.
A weighted least squares approach to the RAS problem was presented by \cite{Rampa2008}.
\cite{Schneider1990}, \cite{Jackson2004}, and \cite{Pavia2009} compared the RAS method with alternative methods formulated as optimization problems.
In some scenarios, the RAS method is the best option, while in others, it can be outperformed.
However, a major advantage of the RAS method over the optimization methods is its computational simplicity.

\subsection{Disaggregation of Input--Output Tables}
\label{sec:litDisagg}

Non-survey methods for estimating regional input--output tables can be categorized as location quotient techniques, commodity balance techniques, and RAS techniques (see \citealp{Harrigan1980, Round1983}).
Several variants of the location quotient method have been proposed and compared by \cite{Schaffer1969}, \cite{Round1978}, \cite{Flegg1995}, \cite{Klijs2016}, and \cite{Flegg2016}.
The commodity balance method was suggested by \cite{Isard1953}.
\cite{Kronenberg2009} presented the cross-hauling adjusted regionalization method, which is closely related to the commodity balance method.
\cite{Tobben2015} modified this method for multi-regional tables.
\cite{Oosterhaven1986} examined the impact of updating inter-regional input--output tables in comparison with single-region tables.

The RAS-based method is not the only method used for disaggregating data and benchmarking.
Many projects which aim to achieve a higher level of detail in the data are a combination of surveyed data, interpolation, and benchmarking.
For example, the FIGARO project is based on ``hard''/surveyed data, gravity models, SUT-RAS, and the 3D variant of GRAS \citep{Remond-Tiedrez2019}.
A similar procedure is used in several other input--output databases.
The EXIOBASE is a detailed multiregional system of SUTs which is based on national SUTs with other national accounts data, the UN Comtrade database, and other data sources \citep{Wood2015}.
The Eora global multi-region input-output database is obtained using the quadratic programming approach and the non-sign-preserving KRAS method \citep{Lenzen2012, Lenzen2013}.
\cite{Geschke2014} compare the construction of the EXIOBASE and the Eora database.
The single-regional input--output tables for the Czech Republic are based on surveyed data (for use, value added, and totals of intermediate use) and the final estimate of the intermediate use per region is a combination of the RAS method and expert estimates \citep{Sixta2016}.

Another problem is the temporal disaggregation of input--output tables.
Dynamic input--output models were studied by, e.g., \cite{Raa1986}, \cite{Aulin-Ahmavaara1990}, and \cite{Ryaboshlyk2006}.
\cite{Avelino2017} disaggregated annual input--output tables along the temporal dimension in order to uncover intra-year seasonality and temporal shocks in the input--output framework.

Input--output tables can also be disaggregated to more detailed industries or sectors.
This was investigated by, e.g., \cite{Katz1981} and \cite{Wolsky1984}.

\section{Methodology}
\label{sec:meth}

\subsection{Input--Output Tables}
\label{sec:methTables}

We start with a description of standard two-dimensional input--output industry-by-industry tables.
Let us assume that the economy is divided into $N$ industries.
An input--output table $X$ consists of non-negative elements $x_{i,j}\geq0$, $i=1,\ldots,N$, $j=1,\ldots,N$, which represent the flow from industry $i$ to industry $j$.
The sum $y^{[1]}_i = \sum_{j=1}^N x_{i,j}$, $i=1,\ldots,N$ represents the total flow from industry $i$ to all industries, while the sum $y^{[2]}_j = \sum_{i=1}^N x_{i,j}$, $j=1,\ldots,N$ represents the total flow from all industries to industry $j$.
The table can also include an additional column representing the final demand and an additional row representing the value added.

Next, we introduce our general notation for multidimensional input--output tables.
Let $D$ denote the number of dimensions and $N_d$ the number of elements in dimension $d$, $d=1,\ldots,D$.
A $D$-dimensional input--output table is then denoted by $\bmni{X}=(x_{n_1,\ldots,n_D}) \in \mathbb{R}_{0}^{N_1 \times \cdots \times N_D}$.
Its totals in the $d$-th dimension are given by
\begin{equation}
\label{eq:totalsDef}
y_{n_1,\ldots,n_{d-1},n_{d+1},\ldots,n_D}^{[d]} = \sum_{n_d=1}^{N_d} x_{n_1,\ldots,n_D},
\end{equation}
for all $n_1,\ldots,n_{d-1},n_{d+1},\ldots,n_D$, i.e., each one is the sum of the elements of $\bmni{X}$ in a single dimension.
The $D-1$-dimensional table of totals in the $d$-th dimension is then denoted by $\bmni{Y}^{[d]}=(y_{n_1,\ldots,n_{d-1},n_{d+1},\ldots,n_D}^{[d]}) \in \mathbb{R}_{+}^{N_1 \times \cdots \times N_{d-1} \times N_{d+1} \times \cdots \times N_D}$.
The sum of all elements of $\bmni{X}$ is given by
\begin{equation}
\label{eq:numberDef}
Z = \sum_{n_{1}=1}^{N_{1}} \cdots \sum_{n_{D}=1}^{N_{D}} x_{n_1,\ldots,n_D}.
\end{equation}
Note that it can be expressed using the totals in the $d$-th dimension as
\begin{equation}
\label{eq:numberAlt}
Z = \sum_{n_{1}=1}^{N_{1}} \cdots \sum_{n_{d-1}=1}^{N_{d-1}} \sum_{n_{d+1}=1}^{N_{d+1}} \cdots \sum_{n_{D}=1}^{N_{D}} y_{n_1,\ldots,n_{d-1},n_{d+1},\ldots,n_D}^{[d]}.
\end{equation}
The notation of standard input--output tables is the special case with $D=2$ and $N_1=N_2=N$.

Our goal is to take an initial $D$-dimensional table $\bmni{X}^{(0)}=(x_{n_1,\ldots,n_D}^{(0)}) \in \mathbb{R}_{0}^{N_1 \times \cdots \times N_D}$ with non-negative elements and find a table $\bmni{X}$ consistent with the given totals $\bmni{Y}^{[1]}, \ldots,\bmni{Y}^{[D]}$ defined in \eqref{eq:totalsDef}.
Note that the totals themselves must be consistent.
Specifically, we require that the given totals in every combination of dimensions $d$ and $e$ (without loss of generality $d<e$) must satisfy
\begin{equation}
\label{eq:totalsCond}
\sum_{n_d=1}^{N_d} y_{n_1,\ldots, n_{e-1}, n_{e+1}, \ldots, n_D}^{[e]} = \sum_{n_e=1}^{N_e} y_{n_1,\ldots,n_{d-1}, n_{d+1}, \ldots, n_D}^{[d]},
\end{equation}
for all $n_1, \ldots, n_{d-1}, n_{d+1},\ldots,n_{e-1},n_{e+1},\ldots,n_D$.
This set of conditions ensures that the totals indeed behave as the sums of elements of $X$.

\subsection{The RAS Method}
\label{sec:methRas}

The \emph{RAS method} is an iterative algorithm which iteratively computes tables $\bmni{X}^{(t)}$,
where $t$ denotes which iteration has been reached.
Each iteration consists of an adjustment in a single dimension $d$.
The first iteration starts with an adjustment in dimension 1, the second iteration concerns dimension 2, and so on.
After the number of dimensions has been reached, the next iteration starts again with an adjustment in dimension 1.
Therefore, the number of iterations can be decomposed as $t = (e - 1) D + d$ where $d=1,\ldots,D$ is the current dimension being adjusted and $e \in \mathbb{N}$.
We can also think of $d$ and $e$ as functions of $t$ given by
\begin{equation}
\begin{aligned}
d(t) &= \left( (t - 1) \bmod D \right) + 1, \\
e(t) &= \left( (t - 1) \div D \right) + 1.\\
\end{aligned}
\end{equation}
In each iteration $t$, the elements of $\bmni{X}^{(t)}$ are given by
\begin{equation}
\label{eq:rasIter}
x_{n_1,\ldots,n_D}^{(t)} = x_{n_1,\ldots,n_D}^{(t-1)} \frac{ y_{n_1,\ldots,n_{d(t)-1},n_{d(t)+1},\ldots,n_D}^{[d(t)]} }{ \sum_{k=1}^{N_{d(t)}} x_{n_1,\ldots,n_{d(t)-1},k,n_{d(t)+1},\ldots,n_D}^{(t-1)} },
\end{equation}
for all $n_1, \ldots, n_D$.
We denote the multipliers of table $\bmni{X}^{(t-1)}$ in dimension $d(t)$ by
\begin{equation}
f_{n_1,\ldots,n_{d(t)-1},n_{d(t)+1},\ldots,n_D}^{[d(t)]} \left( \bmni{X}^{(t-1)} \right) = \frac{ y_{n_1,\ldots,n_{d(t)-1},n_{d(t)+1},\ldots,n_D}^{[d(t)]} }{ \sum_{k=1}^{N_{d(t)}} x_{n_1,\ldots,n_{d(t)-1},k,n_{d(t)+1},\ldots,n_D}^{(t-1)} },
\end{equation}
for all $n_1, \ldots, n_{d(t)-1}, n_{d(t)+1}\ldots,n_D$.
We can then rewrite \eqref{eq:rasIter} as
\begin{equation}
\label{eq:rasRec}
\begin{aligned}
x_{n_1,\ldots,n_D}^{(t)} &= x_{n_1,\ldots,n_D}^{(t-1)} f_{n_1,\ldots,n_{d(t)-1},n_{d(t)+1},\ldots,n_D}^{[d(t)]} \left( \bmni{X}^{(t-1)} \right) \\
&= x_{n_1,\ldots,n_D}^{(0)} \prod_{s=1}^{t} f_{n_1,\ldots,n_{d(s)-1},n_{d(s)+1},\ldots,n_D}^{[d(s)]} \left( \bmni{X}^{(s-1)} \right),
\end{aligned}
\end{equation}
for all $n_1, \ldots, n_D$.
Note that the functions $f_{n_1,\ldots,n_{d(s)-1},n_{d(s)+1},\ldots,n_D}^{[d(s)]}$ depend on intermediate tables $\bmni{X}^{(s-1)}$ and therefore \eqref{eq:rasRec} is in a recursive form.
The RAS solution is then given by
\begin{equation}
\label{eq:rasSolution}
x_{n_1,\ldots,n_D}^{RAS} = \lim_{t \to \infty} x_{n_1,\ldots,n_D}^{(t)},
\end{equation}
for all $n_1, \ldots, n_D$.

For practical computations, the algorithm is stopped when some distance between $\bmni{X}^{(t)}$ and $\bmni{X}^{(t-1)}$ is smaller than a predefined threshold $c$.
The distance can be, e.g., the \emph{Frobenius norm}, giving us the termination condition
\begin{equation}
\sqrt{ \sum_{n_1=1}^{N_1} \cdots \sum_{n_D=1}^{N_D} \left( x_{n_1,\ldots,n_D}^{(t)}-x_{n_1,\ldots,n_D}^{(t-1)} \right)^2 } < c.
\end{equation}

\subsection{The Cross-Entropy Model}
\label{sec:methEntropy}

The \emph{cross-entropy} (CE) model is an optimization problem given by
\begin{equation}
\label{eq:entropyModel}
\begin{aligned}
\text{minimize} \quad & \sum_{n_1=1}^{N_1} \cdots \sum_{n_D=1}^{N_D} x_{n_1,\ldots,n_D} \ln \frac{x_{n_1,\ldots,n_D}}{x_{n_1,\ldots,n_D}^{(0)}}, \\
\text{s.t.} \quad & \sum_{n_d=1}^{N_d} x_{n_1,\ldots,n_D} = y_{n_1, \ldots, n_{d-1}, n_{d+1}, \ldots, n_D}^{[d]} & \quad & \forall \ d \quad \forall \  n_1, \ldots, n_{d-1}, n_{d+1}, \ldots, n_D, \\
& x_{n_1,\ldots,n_D} \geq 0 & \quad & \forall \  n_1, \ldots, n_D.\\
\end{aligned}
\end{equation}
We denote the optimal solution of \eqref{eq:entropyModel} by $x_{n_1,\ldots,n_D}^{CE}$ for all $n_1, \ldots, n_D$.
The term $x_{n_1,\ldots,n_D} \ln x_{n_1,\ldots,n_D}$ is defined at zero as its limiting value
\begin{equation}
\lim_{x_{n_1,\ldots,n_D} \to 0} x_{n_1,\ldots,n_D} \ln x_{n_1,\ldots,n_D} = 0.
\end{equation}
When $x_{n_1,\ldots,n_D}^{(0)}$ is zero, the optimal $x_{n_1,\ldots,n_D}^{CE}$ must clearly also be zero, as 
\begin{equation}
\lim_{\substack{x_{n_1,\ldots,n_D} \to 0, \\ x_{n_1,\ldots,n_D}^{(0)} \to 0}} x_{n_1,\ldots,n_D} \ln \frac{1}{x_{n_1,\ldots,n_D}^{(0)}} = 0, \qquad
\lim_{x_{n_1,\ldots,n_D}^{(0)} \to 0} \ln \frac{1}{x_{n_1,\ldots,n_D}^{(0)}} = \infty.
\end{equation}
We therefore restrict ourselves to $x_{n_1,\ldots,n_D} = 0$ when $x_{n_1,\ldots,n_D}^{(0)} = 0$.

Let us briefly present some properties of the model.
The rank of the matrix of conditions is
\begin{equation}
R(N_1, N_2, \ldots, N_D) = \sum_{i=1}^{D} (-1)^{i+1} \sum_{1 \leq j_1 < j_2 < \cdots < j_{D-i} \leq D} N_{j_1} N_{j_2} \cdots N_{j_{D-i}}.
\end{equation}
Note that for all $N_d > 1$, $d=1,\ldots,D$, we have $N_1 N_2 \cdots N_D > R(N_1, N_2, \ldots, N_D)$, i.e., the number of variables is strictly greater than the rank of the matrix of conditions.
This is a necessary condition for the system of linear equations to be consistent.
It also implies that the problem has either no feasible solution (in the case of inconsistent constraints) or an infinite number of solutions.
For example, in the three-dimensional case with $N_1=2$, $N_2=3$ and $N_3=4$, there are $2 \times 3 \times 4 = 24$ variables and $3 \times 4 + 2 \times 4 + 2 \times 3 = 26$ constraints but their rank is only $3 \times 4 + 2 \times 4 + 2 \times 3 - 2 - 3 - 4 + 1 = 18$.
The objective function is bounded from below, as shown in Lemma \ref{th:bound}, and is strictly convex, as shown in Lemma \ref{th:convex}.
In both lemmas, we omit zero values of $x_{n_1,\ldots,n_D}^{(0)}$ as in that case, we do not consider $x_{n_1,\ldots,n_D}$ to be a variable but a constant $x_{n_1,\ldots,n_D} = 0$.

\begin{lemma}
\label{th:bound}
Let $\bmni{X}^{(0)}=(x_{n_1,\ldots,n_D}^{(0)}) \in \mathbb{R}_{+}^{N_1 \times \cdots \times N_D}$ be a $D$-dimensional table with positive elements.
The function $f(\bmni{X}) = \sum_{n_1=1}^{N_1} \cdots \sum_{n_D=1}^{N_D} x_{n_1,\ldots,n_D} \ln ( x_{n_1,\ldots,n_D} / x_{n_1,\ldots,n_D}^{(0)} )$ is then bounded from below by $f(\bmni{X}) \geq Z - Z^{(0)}$, where $Z$ denotes the sum of all elements of $\bmni{X}$ and $Z^{(0)}$ of $\bmni{X}^{(0)}$.
\end{lemma}

\begin{proof}
This proof follows the proof of the Gibbs inequality and uses the inequality $\ln x \leq x - 1$ for $x>0$.
For all $x_{n_1,\ldots,n_D} > 0$, we have
\begin{equation}
\label{eq:entropyGibbs}
\begin{aligned}
\sum_{n_1=1}^{N_1} \cdots \sum_{n_D=1}^{N_D} x_{n_1,\ldots,n_D} \ln \frac{x_{n_1,\ldots,n_D}}{x_{n_1,\ldots,n_D}^{(0)}} &= - \sum_{n_1=1}^{N_1} \cdots \sum_{n_D=1}^{N_D} x_{n_1,\ldots,n_D} \ln \frac{x_{n_1,\ldots,n_D}^{(0)}}{x_{n_1,\ldots,n_D}} \\
&\geq  - \sum_{n_1=1}^{N_1} \cdots \sum_{n_D=1}^{N_D} x_{n_1,\ldots,n_D} \left( \frac{x_{n_1,\ldots,n_D}^{(0)}}{x_{n_1,\ldots,n_D}} - 1 \right) \\
&=  - \sum_{n_1=1}^{N_1} \cdots \sum_{n_D=1}^{N_D} \left( x_{n_1,\ldots,n_D}^{(0)} - x_{n_1,\ldots,n_D} \right) \\
&= \sum_{n_1=1}^{N_1} \cdots \sum_{n_D=1}^{N_D} x_{n_1,\ldots,n_D} - \sum_{n_1=1}^{N_1} \cdots \sum_{n_D=1}^{N_D} x_{n_1,\ldots,n_D}^{(0)}  \\
&= Z - Z^{(0)}.\\
\end{aligned}
\end{equation}
For $x_{n_1,\ldots,n_D} = 0$, we have $x_{n_1,\ldots,n_D} \ln x_{n_1,\ldots,n_D} / x_{n_1,\ldots,n_D}^{(0)} = 0$ and \eqref{eq:entropyGibbs} also holds.
The objective function is therefore bounded from below by the difference between the sum of all elements of $\bmni{X}$ and $\bmni{X}^{(0)}$.
\end{proof}

\begin{lemma}
\label{th:convex}
Let $\bmni{X}^{(0)}=(x_{n_1,\ldots,n_D}^{(0)}) \in \mathbb{R}_{+}^{N_1 \times \cdots \times N_D}$ be a $D$-dimensional table with positive elements.
The function $f(\bmni{X}) = \sum_{n_1=1}^{N_1} \cdots \sum_{n_D=1}^{N_D} x_{n_1,\ldots,n_D} \ln (x_{n_1,\ldots,n_D} / x_{n_1,\ldots,n_D}^{(0)})$ is then strictly convex.
\end{lemma}

\begin{proof}
First, we focus on the one-dimensional case of the function
\begin{equation}
g(x_{n_1,\ldots,n_D}) =  x_{n_1,\ldots,n_D} \ln \frac{x_{n_1,\ldots,n_D}}{x_{n_1,\ldots,n_D}^{(0)}}.
\end{equation}
For $x_{n_1,\ldots,n_D} > 0$, it is twice differentiable with strictly positive second derivative
\begin{equation}
\frac{\partial^2 g}{\partial x_{n_1,\ldots,n_D}^2} = \frac{1}{x_{n_1,\ldots,n_D}} > 0.
\end{equation}
Therefore, it is convex for $x_{n_1,\ldots,n_D} > 0$.
At $x_{n_1,\ldots,n_D} = 0$, however, it does not have a second derivative.
To extend the range to $x_{n_1,\ldots,n_D} \geq 0$, we will use the definition of strict convexity.
We will show that 
\begin{equation}
g \left( \lambda x_{n_1,\ldots,n_D}' + (1-\lambda) x_{n_1,\ldots,n_D}'' \right) < \lambda g \left( x_{n_1,\ldots,n_D}' \right) + (1-\lambda) g \left( x_{n_1,\ldots,n_D}'' \right)
\end{equation}
for all $x_{n_1,\ldots,n_D}' > 0$, $x_{n_1,\ldots,n_D}'' = 0$ and all $\lambda \in (0,1)$.
As $g(x_{n_1,\ldots,n_D}'') = 0$, we have
\begin{equation}
\begin{aligned}
g \left( \lambda x_{n_1,\ldots,n_D}' \right) &< \lambda g \left( x_{n_1,\ldots,n_D}' \right) \\
\lambda x_{n_1,\ldots,n_D}' \ln \frac{\lambda x_{n_1,\ldots,n_D}'}{x_{n_1,\ldots,n_D}^{(0)}} &< \lambda x_{n_1,\ldots,n_D}' \ln \frac{x_{n_1,\ldots,n_D}'}{x_{n_1,\ldots,n_D}^{(0)}} \\
\ln \lambda + \ln x_{n_1,\ldots,n_D}' - \ln x_{n_1,\ldots,n_D}^{(0)} &< \ln x_{n_1,\ldots,n_D}' - \ln x_{n_1,\ldots,n_D}^{(0)} \\
\ln \lambda &< 0.\\
\end{aligned}
\end{equation}
The last inequality holds since $\lambda \in (0,1)$.
Therefore, $g(x_{n_1,\ldots,n_D})$ is convex for $x_{n_1,\ldots,n_D} \geq 0$.
Finally,
\begin{equation}
f(\bmni{X}) = \sum_{n_1=1}^{N_1} \cdots \sum_{n_D=1}^{N_D} g(x_{n_1,\ldots,n_D})
\end{equation}
is strictly convex since it is a sum of strictly convex functions.
\end{proof}

\subsection{Relation of the RAS Method to the Cross-Entropy Model}
\label{sec:methRelation}

The solution of the RAS method converges to the solution of the cross-entropy model.
This is an important result, since it is possible to quite efficiently compute the solution to the non-linear optimization problem of the cross-entropy by the iterative RAS method.
We demonstrate the equivalence between the two methods in the following theorem.

\begin{theorem}
\label{th:relation}
Suppose $\bmni{X}^{(0)}=\left(x_{n_1,\ldots,n_D}^{(0)}\right) \in \mathbb{R}_{0}^{N_1 \times \cdots \times N_D}$ is a $D$-dimensional table with non-negative elements and $\bmni{Y}^{[d]}=(y_{n_1,\ldots,n_{d-1},n_{d+1},\ldots,n_D}^{[d]}) \in \mathbb{R}_{+}^{N_1 \times \cdots \times N_{d-1} \times N_{d+1} \times \cdots \times N_D}$, $d=1,\ldots,D$ are total tables with positive elements satisfying \eqref{eq:totalsCond}.
Suppose the RAS procedure given by \eqref{eq:rasSolution} converges to $\bmni{X}^{RAS}=(z^{RAS}_{n_1,\ldots,n_D}) \in \mathbb{R}_{0}^{N_1 \times \cdots \times N_D}$ satisfying \eqref{eq:totalsDef} and suppose the cross-entropy optimization problem \eqref{eq:entropyModel} has unique solution $\bmni{X}^{CE}=(z^{CE}_{n_1,\ldots,n_D}) \in \mathbb{R}_{0}^{N_1 \times \cdots \times N_D}$.
The RAS solution $\bmni{X}^{RAS}$ is then the same as the cross-entropy solution $\bmni{X}^{CE}$.
\end{theorem}

\begin{proof}
We base the proof on the two-dimensional case of \cite{Lemelin2013}.
First, let us find expressions for the multipliers of the RAS method.
We separate the products of $f_{n_1,\ldots,n_{d(s)-1},n_{d(s)+1},\ldots,n_D}^{[d(s)]}$ in \eqref{eq:rasSolution} according to their adjusting dimension.
The multipliers in dimension $d$ are
\begin{equation}
\theta_{n_1,\ldots,n_{d-1},n_{d+1},\ldots,n_D}^{[d]} = \lim_{t \to \infty} \prod_{s: s \leq t \wedge d(s)=d} f_{n_1,\ldots,n_{d-1},n_{d+1},\ldots,n_D}^{[d]} \left( \bmni{X}^{(s-1)} \right),
\end{equation}
for all $n_1, \ldots, n_{d-1}, n_{d+1}\ldots,n_D$.
The RAS solution \eqref{eq:rasSolution} can then be rewritten as
\begin{equation}
\label{eq:rasMultip}
x_{n_1,\ldots,n_D}^{RAS} = x_{n_1,\ldots,n_D}^{(0)} \prod_{d=1}^D \theta_{n_1,\ldots,n_{d-1},n_{d+1},\ldots,n_D}^{[d]},
\end{equation}
for all $n_1, \ldots, n_D$.
We sum equations \eqref{eq:rasMultip} in dimension $d$ and get
\begin{equation}
\label{eq:rasMultipSum}
\sum_{n_d=1}^{N_d} x_{n_1,\ldots,n_D}^{RAS} = \sum_{n_d=1}^{N_d} x_{n_1,\ldots,n_D}^{(0)} \prod_{d=1}^D \theta_{n_1,\ldots,n_{d-1},n_{d+1},\ldots,n_D}^{[d]},
\end{equation}
for all $n_1, \ldots, n_{d-1}, n_{d+1}\ldots,n_D$.
The left side of \eqref{eq:rasMultipSum} is equal to $y_{n_1,\ldots,n_{d-1},n_{d+1},\ldots,n_D}^{[d]}$ according to \eqref{eq:totalsDef} and we can rewrite \eqref{eq:rasMultipSum} as
\begin{equation}
\label{eq:rasFinal}
\theta_{n_1,\ldots,n_{d-1},n_{d+1},\ldots,n_D}^{[d]} = y_{n_1,\ldots,n_{d-1},n_{d+1},\ldots,n_D}^{[d]} \left( \sum_{n_d=1}^{N_d} x_{n_1,\ldots,n_D}^{(0)} \prod_{e \neq d} \theta_{n_1,\ldots,n_{e-1},n_{e+1},\ldots,n_D}^{[e]} \right)^{-1},
\end{equation}
for all $n_1, \ldots, n_{d-1}, n_{d+1}\ldots,n_D$.
Second, let us express the cross-entropy solution in a similar fashion.
The Lagrangian of \eqref{eq:entropyModel} is given by
\begin{equation}
\label{eq:lagrang}
\begin{aligned}
L &= \sum_{n_1=1}^{N_1} \cdots \sum_{n_D=1}^{N_D} x_{n_1,\ldots,n_D} \ln \frac{x_{n_1,\ldots,n_D}}{x_{n_1,\ldots,n_D}^{(0)}} \\
& \ + \sum_{d=1}^D \left( \sum_{n_1=1}^{N_1} \cdots \sum_{n_D=1}^{N_D} \lambda_{n_1,\ldots,n_{d-1},n_{d+1},\ldots,n_D}^{[d]} \left( x_{n_1,\ldots,n_D} - \frac{y_{n_1, \ldots, n_{d-1}, n_{d+1}, \ldots, n_D}^{[d]}}{N_d} \right) \right).
\end{aligned}
\end{equation}
The first derivatives of the Lagrangian with respect to the multipliers are
\begin{equation}
\label{eq:lagrangDerMulti}
\frac{\partial L}{\partial \lambda_{n_1,\ldots,n_{d-1},n_{d+1},\ldots,n_D}^{[d]}} = \sum_{n_d=1}^{N_d} x_{n_1,\ldots,n_D} - y_{n_1, \ldots, n_{d-1}, n_{d+1}, \ldots, n_D}^{[d]},
\end{equation}
for $d=1,\ldots,D$ and for all $n_1, \ldots, n_{d-1}, n_{d+1}\ldots,n_D$.
The first derivatives of the Lagrangian with respect to the original variables are
\begin{equation}
\label{eq:lagrangDerVar}
\frac{\partial L}{\partial x_{n_1,\ldots,n_D}} = 1 + \ln x_{n_1,\ldots,n_D} - \ln x_{n_1,\ldots,n_D}^{(0)} + \sum_{d=1}^D \lambda_{n_1,\ldots,n_{d-1},n_{d+1},\ldots,n_D}^{[d]},
\end{equation}
for all $n_1, \ldots, n_D$.
The first derivatives must be equal to 0 for the optimal solution $x_{n_1,\ldots, n_D}^{CE}$ for all $n_1, \ldots, n_D$.
As the objective function is strictly convex, by Lemma \ref{th:convex}, the first-order conditions are sufficient and have a unique solution.
Setting \eqref{eq:lagrangDerVar} to 0 and using the exponential function we have
\begin{equation}
x_{n_1,\ldots, n_D}^{CE} = x_{n_1,\ldots, n_D}^{(0)} e^{ -1 - \sum_{d=1}^{D} \lambda_{n_1,\ldots,n_{d-1},n_{d+1},\ldots,n_D}^{[d]} },
\end{equation}
for all $n_1, \ldots, n_D$.
In each dimension $d$, we use the substitution
\begin{equation}
\varphi_{n_1,\ldots,n_{d-1},n_{d+1},\ldots,n_D}^{[d]} = \left\{
\begin{array}{ll}
e^{ - 1 - \lambda_{n_1,\ldots,n_{d-1},n_{d+1},\ldots,n_D}^{[d]} } & \quad \text{for } d = 1, \\
e^{ - \lambda_{n_1,\ldots,n_{d-1},n_{d+1},\ldots,n_D}^{[d]} } & \quad \text{for } d > 1, \\
\end{array}
\right.
\end{equation}
for all $n_1, \ldots, n_{d-1}, n_{d+1}\ldots,n_D$.
The solution of \eqref{eq:entropyModel} is then
\begin{equation}
\label{eq:entropyMultip}
x_{n_1,\ldots,n_D}^{CE} = x_{n_1,\ldots,n_D}^{(0)} \prod_{d=1}^D \varphi_{n_1,\ldots,n_{d-1},n_{d+1},\ldots,n_D}^{[d]},
\end{equation}
for all $n_1, \ldots, n_D$.
We sum equations \eqref{eq:entropyMultip} in dimension $d$ and get
\begin{equation}
\label{eq:entropyMultipSum}
\sum_{n_d=1}^{N_d} x_{n_1,\ldots,n_D}^{CE} = \sum_{n_d=1}^{N_d} x_{n_1,\ldots,n_D}^{(0)} \prod_{d=1}^D \varphi_{n_1,\ldots,n_{d-1},n_{d+1},\ldots,n_D}^{[d]},
\end{equation}
for all $n_1, \ldots, n_{d-1}, n_{d+1}\ldots,n_D$.
We set \eqref{eq:lagrangDerMulti} to 0 and the left side of \eqref{eq:entropyMultipSum} is therefore equal to $y_{n_1,\ldots,n_{d-1},n_{d+1},\ldots,n_D}^{[d]}$.
We can rewrite \eqref{eq:entropyMultipSum} as
\begin{equation}
\label{eq:entropyFinal}
\varphi_{n_1,\ldots,n_{d-1},n_{d+1},\ldots,n_D}^{[d]} = y_{n_1,\ldots,n_{d-1},n_{d+1},\ldots,n_D}^{[d]} \left( \sum_{n_d=1}^{N_d} x_{n_1,\ldots,n_D}^{(0)} \prod_{e \neq d} \varphi_{n_1,\ldots,n_{e-1},n_{e+1},\ldots,n_D}^{[e]} \right)^{-1},
\end{equation}
for all $n_1, \ldots, n_{d-1}, n_{d+1}\ldots,n_D$.
Finally, let us compare the RAS solution and the cross-entropy solution.
According to \eqref{eq:rasMultip} and \eqref{eq:entropyMultip}, both solutions are in the same multiplicative form, consisting of $x_{n_1,\ldots,n_D}^{(0)}$ and multipliers corresponding to each dimension.
According to \eqref{eq:rasFinal} and \eqref{eq:entropyFinal}, these multipliers must satisfy exactly the same system of equations.
The RAS equations \eqref{eq:rasMultip} with \eqref{eq:rasFinal} have a unique solution and the CE equations  \eqref{eq:entropyMultip} with \eqref{eq:entropyFinal} also have a unique solution.
Therefore, the multipliers have the same values and $x_{n_1,\ldots, n_D}^{RAS} = x_{n_1,\ldots, n_D}^{CE}$ for all $n_1, \ldots, n_D$.
\end{proof}

Note that the formulated cross-entropy theorem is related to the pure general form of the optimization problem and does not take into account any other restrictions which input--output theorists/practitioners use and need.

\section{Applications}
\label{sec:app}

\subsection{Single-Regional Input--Output Tables}
\label{sec:appSingle}

We use the multidimensional RAS method to estimate regional input--output tables for the Czech Republic.
We use industry-by-industry tables with a total of 82 industries.
We have a national input--output table and row and column totals for regional tables for the year 2011.
Our goal is to estimate the input--output table for each region.
The Czech Republic has 14 regions: Prague (R01), Central Bohemia (R02), South Bohemia (R03), Plze\v{n} (R04), Karlovy Vary (R05), \'{U}st\'{i} nad Labem (R06), Liberec (R07), Hradec Kr\'{a}lov\'{e} (R08), Pardubice (R09), Vyso\v{c}ina (R10), South Moravia (R11), Olomouc (R12), Zl\'{i}n (R13), and Moravia-Silesia (R14).
Region R01 is the capital city.
The original regional input--output tables were constructed according to the method of \cite{Sixta2015} and \cite{Sixta2016}.
Note that in all our applications, the tables do not contain any negative values.
The RIOT/SIOT tables were pre-processed by the Czech Statistical Office and there are no negative values in the intermediate use.

\begin{table}
\begin{center}
\caption{The Frobenius norm of the difference between the benchmark tables and the two-dimensional and multidimensional estimates.}
\label{tab:comp}
\begin{tabular}{llrrr}
\toprule
Tables                & Transformation     & 2D RAS & 3D RAS & No.
of elements\\
\midrule
Single-regional       & Total sum          &  \num{11610.10} &      \num{0.00} & $82 \times 82$ \\
Single-regional       & Input--output table &   \num{4798.76} &     \num{94.75} & $14 \times 82 \times 82$ \\
Single-regional       & Leontief inverse   &      \num{0.59} &      \num{0.44} & $14 \times 82 \times 82$ \\
Inter-regional        & Isard's model      &  \num{53696.09} &  \num{50346.88} & $14 \times 82 \times 14 \times 82$ \\
Quarterly             & Total sum          & \num{723175.62} &      \num{0.00} & $21 \times 42 \times 42$ \\
Quarterly             & Value added        & \num{235836.61} & \num{106483.73} & $ 4 \times 21 \times 41$ \\
Domestic/imported     & Total sum          &  \num{48661.00} &      \num{0.00} & $ 4 \times 82 \times 82$ \\
\bottomrule
\end{tabular}
\end{center}
\end{table}

\begin{figure}
\centering
\includegraphics[width=10cm]{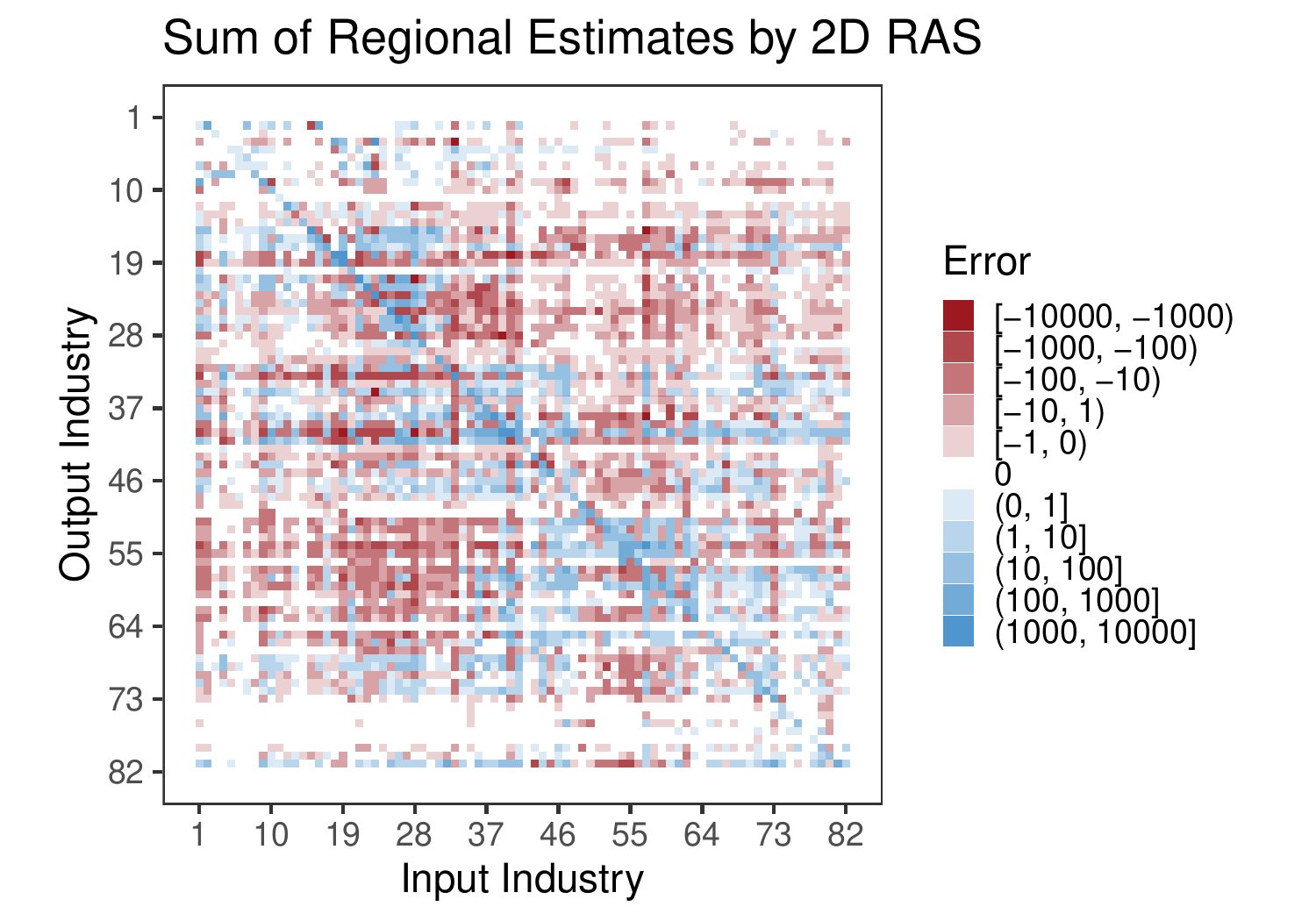} 
\caption{The error of the national input--output table of the Czech Republic in 2011 and the sum of two-dimensional regional estimates.}
\label{fig:regionalTotal}
\end{figure}

We can estimate regional tables by the two-dimensional RAS method using their row and column sums as well as the structure of the national table.
However, this approach does not ensure that all regional tables estimated in this way add up to the national table.
First, we compare the sum of the regional tables estimated by the two-dimensional RAS algorithm and the national table.
 Figure \ref{fig:regionalTotal} shows the errors of the sum of regional tables estimated by the two-dimensional RAS method.
We can see that for some elements, the error can be quite high: the largest positive error is 4866.19 (corresponding to a relative error of 5.33 percent) while the largest negative error is -2418.34 (corresponding to a relative error of -4.98 percent).
If we use the multidimensional RAS method instead, this error is zero by definition.
Figure \ref{fig:regionalTotal} demonstrates that the violation in the third dimension is substantial when using the two-dimensional RAS method, which motivates us to obtain consistent estimates using the multidimensional RAS method.

\begin{figure}
\centering
\includegraphics[width=14cm]{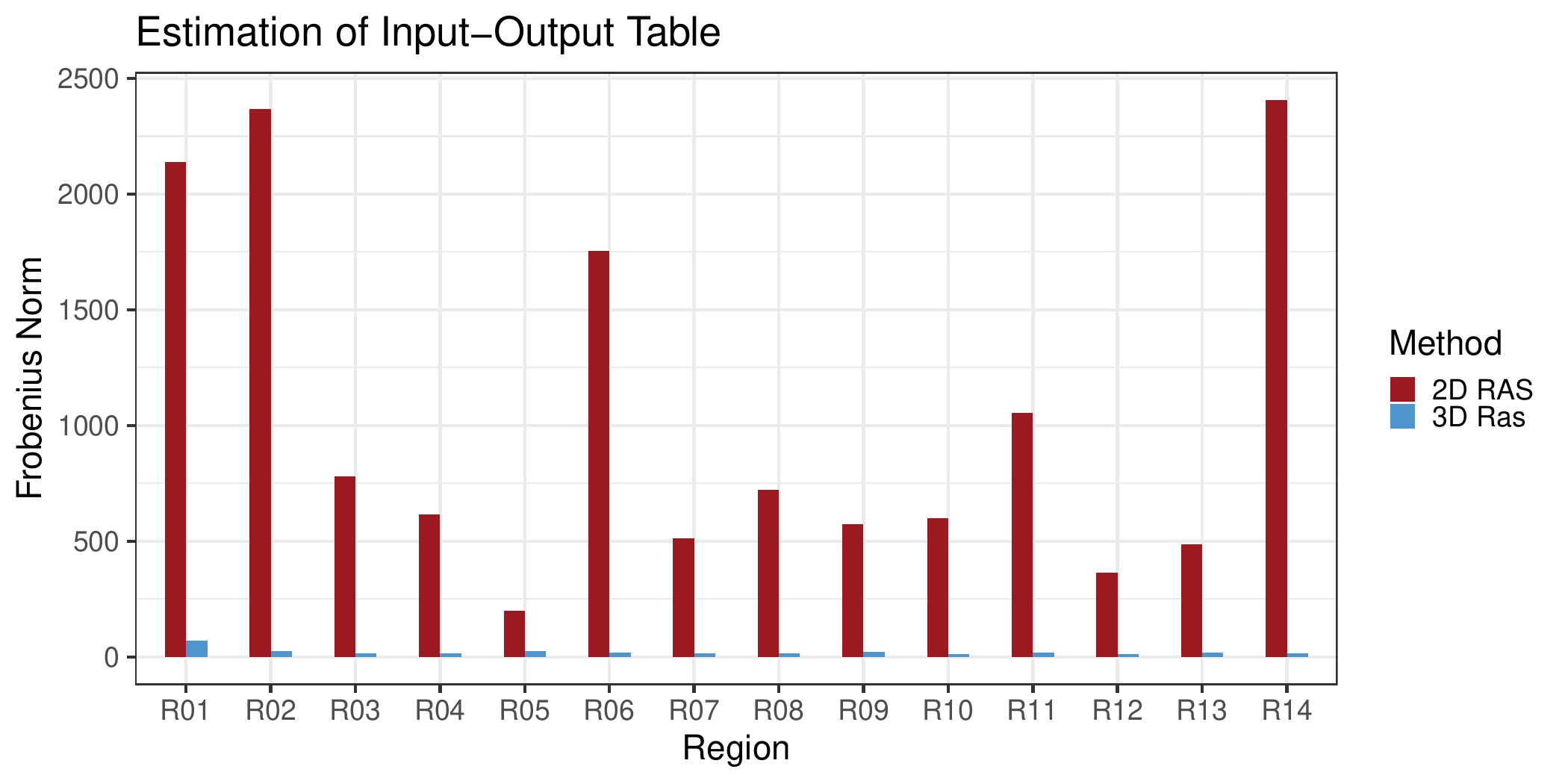} 
\caption{The Frobenius norm of the difference between the benchmark regional input--output tables of the Czech Republic in 2011 and two-dimensional and multidimensional estimates.}
\label{fig:regionalTable}
\end{figure}

We compare the two-dimensional and multidimensional estimates with the regional tables estimated by \cite{Sixta2016}, which we refer to as the benchmark tables.
Their method is based on the model approach and extrapolation.
For example, the total output is a sum of surveyed and estimated items.
The benchmark tables are published on the level of two digits of the CPA classification and follow the method of ESA 2010.
To compare these tables, we use the Frobenius norm.
The results for each region are shown in Figure \ref{fig:regionalTable}.
The total Frobenius norm is shown in Table \ref{tab:comp} for both methods.
We can see that the multidimensional estimates are much closer to the benchmark tables than are the two-dimensional estimates.
The differences between the benchmark tables and the two-dimensional RAS estimates are caused mainly because the two-dimensional RAS does not respect the sum of cells.
The Frobenius norm for the multidimensional estimates varies between 12.38 and 69.44.
The highest deviation of the multidimensional estimate from the benchmark table is for region R01.
This is not surprising, since R01 is the capital city, Prague, and its economic structure differs from the rest of the country.
The multidimensional RAS does not therefore fit the regional tables perfectly, but does fits them significantly better.
However, the benchmark tables, by which we evaluate the performance of these methods, are partially surveyed and partially interpolated from the national table \citep{Sixta2016, Safr2017}.
According to this method, the intermediate matrices are pre-estimated by the two-dimensional RAS and then corrected by additional data sources and expert estimates.
So the additivity is achieved by manual corrections, but the structures themselves are obtained by the two-dimensional RAS.
Consequently, the main source of error for the two-dimensional RAS is the unachieved additivity.

\begin{figure}
\centering
\includegraphics[width=14cm]{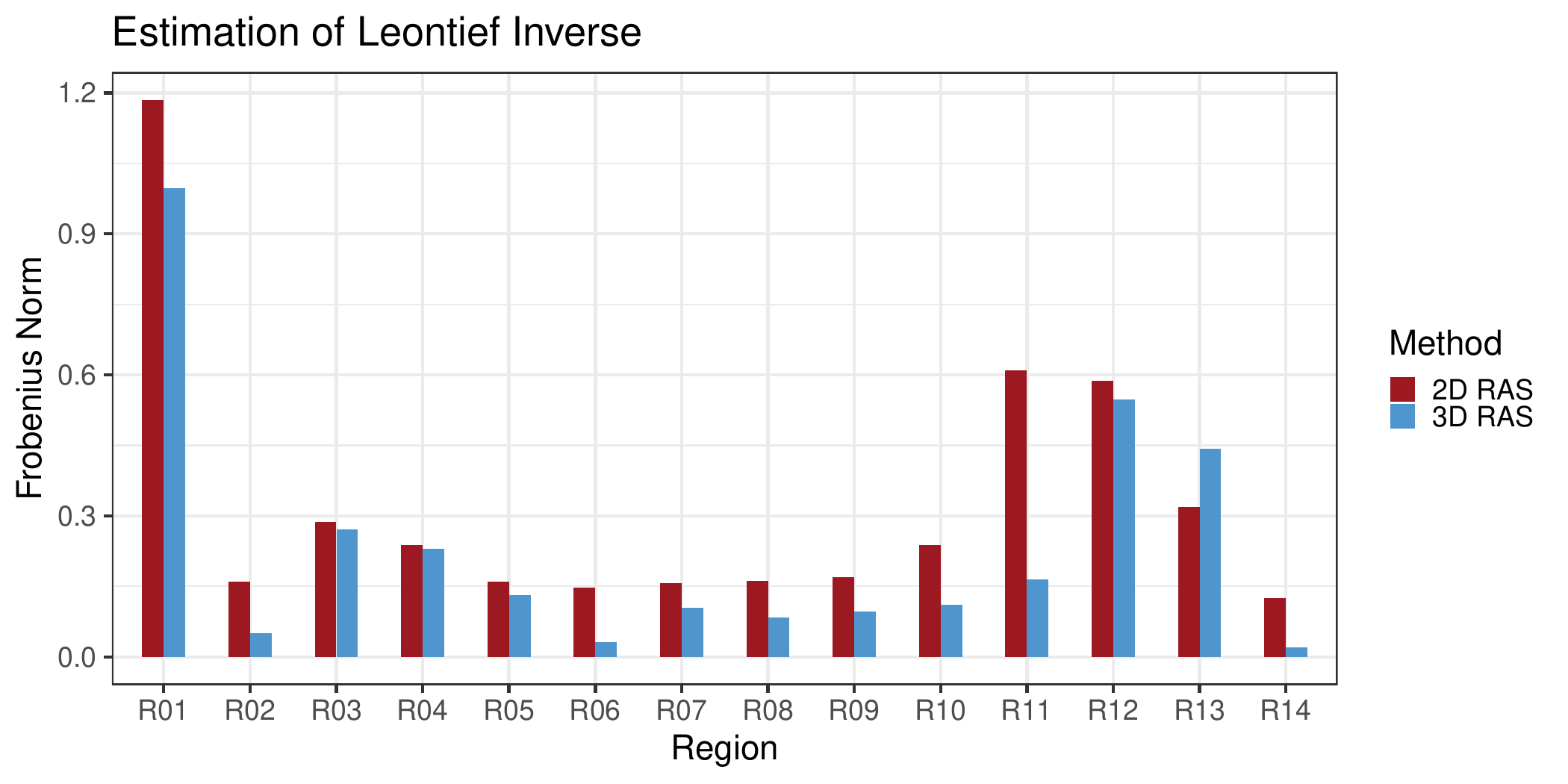} 
\caption{The Frobenius norm of the difference between the Leontief inverse of the benchmark regional input--output tables of the Czech Republic in 2011 and the two-dimensional and multidimensional estimates.}
\label{fig:regionalLeontief}
\end{figure}

Next, we compare the Leontief inverses of the estimates and the benchmark tables.
The Leontief inverse, or the total requirements matrix, is defined by $\bmni{L} = (\bmni{I}-\bmni{A})^{-1}$, where $\bmni{I}$ is the identity matrix, $\bmni{A}=\left( a_{n_1,n_2} \right)_{n_1=1,n_2=1}^{N_1,N_2}$ is the matrix of technical coefficients given by $a_{n_1,n_2}=x_{n_1,n_2} /  r_{n_2}$, and $\bmni{r} = (r_1, \ldots, r_{N_2})'$ is the total production.
This can give us some idea about the predictive ability of the estimations, as the Leontief inverse is often used to predict the total production $\bmni{r}$ from the final consumption $\bmni{f}$ using the equation $\bmni{r} = \bmni{L} \bmni{f}$.
Again, we compare the matrices using the Frobenius norm.
In Figure \ref{fig:regionalLeontief}, we can see that the Leontief inverses of the multidimensional RAS estimates are closer to those of the benchmark tables than are the two-dimensional estimates in all regions, except the Zl\'{i}n region (R13).
However, the difference is not as significant as in the case of the original non-inverted matrices.

Note that, in the presented application, the multidimensional RAS method outperforms the two-dimensional approach not only in terms of average errors but also in terms of extreme errors.
The highest absolute error between the estimated and the benchmark tables is \num{1164.46} for the two-dimensional RAS and \num{55.14} for the multidimensional RAS.
In the case of the Leontief inverse, the highest absolute error is \num{1.10} for the two-dimensional RAS and \num{0.94} for the multidimensional RAS.

\subsection{Inter-Regional Input--Output Tables}
\label{sec:appInter}

As another application, we use the Isard's inter-regional input--output model (IRIO) of \cite{Isard1960}.
The Isard's model extends the single regional model by the values of the inter-regional linkages between each product/industry.
These values are incorporated into one big technical coefficients matrix, final use, and value added.
In this way, the model takes into account the spillover and feedback effects caused by inter-regional linkages.
These models are mainly used to analyze the local effects of national economic policies.
They are also used for the analysis of environmental issues in the context of the individual sectors in the Czech Republic.
Finally, they form the data basis for more advanced models, such as DSGE, discussed in \cite{Bouakez2009} and \cite{Bouakez2014}.
The advantage of the Isard's IRIO model is the fact that this approach allows us to analyze the so-called backward relations in the model.
Given the complexity of this approach, we only summarize the basic facts.
The IRIO model is a disaggregated national model which allows analyzing effects at both the individual industry level as well as the regional unit level.
It can be used in particular to analyze the impact of economic policy.

The main strength of the Isard's model lies in its detailed level of information.
The regions are fully connected and it allows studying and evaluating the linkages and interactions between each region's industry and those of other regions and industries.
In previous models, the inter-regional flow was mainly represented by row/columns that describe only the flow from/to this region.
These models had the information to which regions is the production exported (or from which regions it is imported) but without exact details about the industry.
They did not allow comprehensively studying the forward/backward effects across the regions.

Using the method of \cite{Miller2009} and derived procedure for the inter-regional model based on the Czech Republic data of \cite{Safr2016a}, we construct the inter-regional model for the Czech Republic.
The data used in Section \ref{sec:appSingle} do not contain the detailed allocation of inter-regional production flows.
In this section, the data are constructed according to \cite{Safr2016}.
The model is built entirely on the basis of data sources.
We also construct two additional models.
In these models we assume that we do not know the inter-regional tables describing the use of the industry imports (for each pair of regions).
This missing data source is used for the allocation of inter-regional production flows into the intermediate consumption in other regions.
Then, using the national table of regional flows, we estimate inter-regional matrices using the two-dimensional RAS method and the multidimensional RAS method as well.
For the RAS methods, we therefore assume that we know only the national input--output table (i.e., the sum of the inter-regional input--output tables for the use of imports) and the total flow between the regions for each industry (i.e., the row and column totals of the inter-regional input--output tables).
The two-dimensional RAS method considers $14 \times 14$ independent inter-regional matrices, each with constraints on 82 row totals and 82 column totals.
The multidimensional RAS additionally considers a third dimension consisting of constraints on the sum of all $14 \times 14$ inter-regional matrices.

\begin{figure}
\centering
\includegraphics[width=14cm]{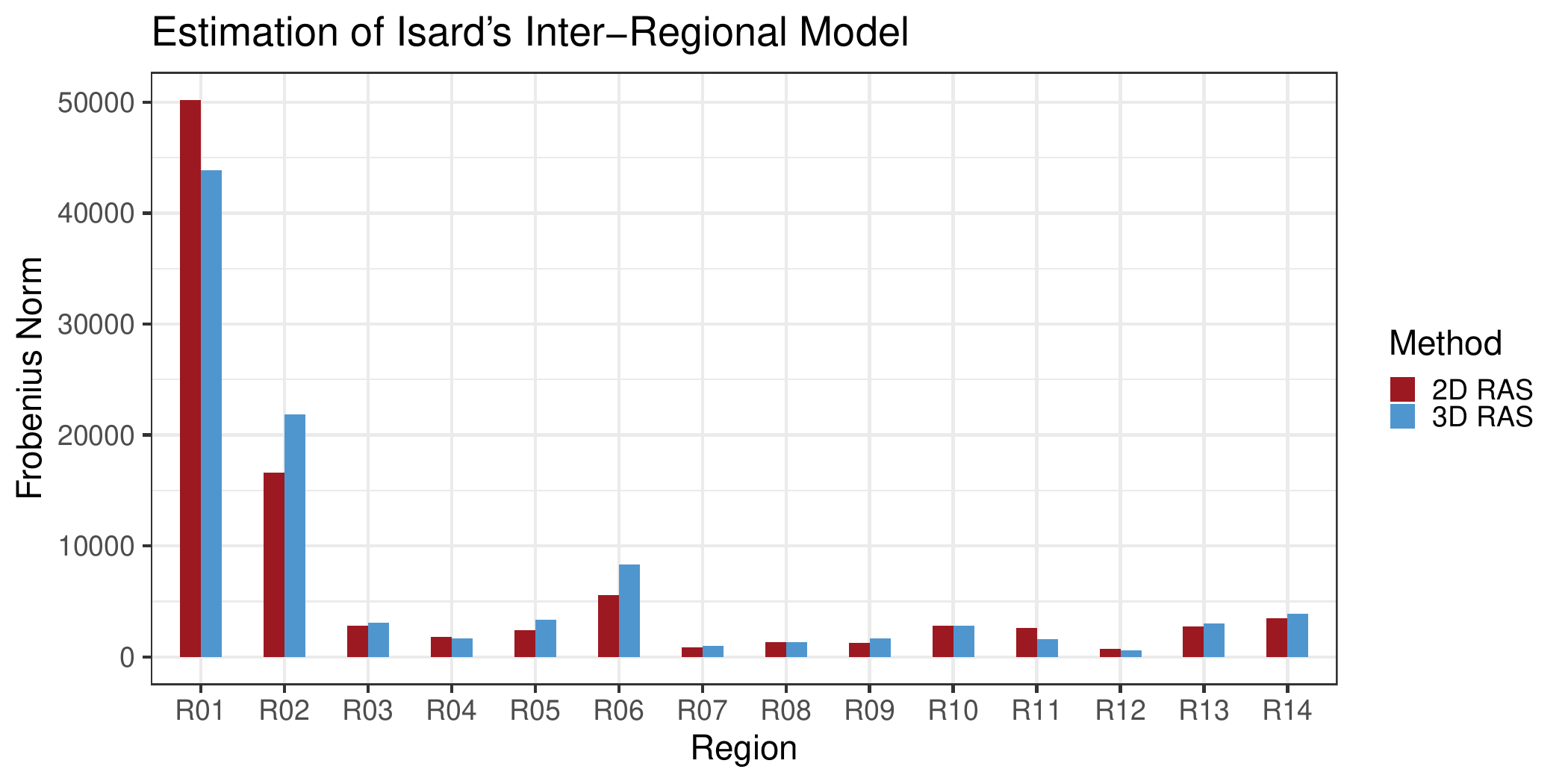} 
\caption{The Frobenius norm of the difference between the benchmark tables in the Isard's inter-regional input--output model of the Czech Republic in 2011 and the two-dimensional and multidimensional estimates.}
\label{fig:regionalIsard}
\end{figure}

The evaluation of the results is performed on the total intermediate consumption matrix for all regions and industries (14 regions each with 82 industries times 14 regions each with 82 industries).
As in the previous sections, we calculate the Frobenius norm.
As we can see in Table \ref{tab:comp}, the multidimensional RAS shows less difference from the model constructed on the basis of benchmark tables.
The two-dimensional RAS approach overestimates by 6.65 percent compared to the multidimensional RAS.
However, this overall statistic can be distorted by a few extreme values.
For a more detailed look, we decompose the Frobenius norm according to the individual regions in Figure \ref{fig:regionalIsard}.
The multidimensional RAS is superior in region R01 while the two-dimensional RAS is superior in regions R02 and R06.
For the other regions, the methods perform quite similarly.
Based on Figure \ref{fig:regionalIsard}, we therefore can not decide unambiguously that one method is systematically better or worse than the other for the individual regions.

\subsection{Quarterly Input--Output Tables}
\label{sec:appQuarterly}

In a similar fashion as in Section \ref{sec:appSingle}, we analyze the difference between the multidimensional RAS method and the two-dimensional RAS method.
This time, the third dimension represents division into four quarters (Q1--Q4).
We now work with tables with 41 industries.
Compared to Sections \ref{sec:appSingle} and \ref{sec:appInter}, the industries are aggregated since the source data, from which the tables are constructed, are only available at this level of division.
We also include the value added and the final demand in our analysis, resulting in matrices with 42 rows and columns.
We have the quarterly total production and value added available for the years 1995--2015.
We also have annual tables available for the years 1995, 2000, 2005, 2010, and 2013.
We estimate the remaining annual tables by the two-dimensional RAS method.

This application of the multidimensional RAS method to quarterly data should illustrate the differences from the two-dimensional approach.
However, we should note that the multidimensional RAS method does not ensure that the estimated tables follow the pure logic behind the data.
For example, in many industries, we can expect the structure to be stable throughout the year (e.g., metallurgy and mining).
On the other hand, some industries may have seasonal effects in the structure (e.g., agriculture and construction).

First, we estimate the quarterly matrices by the two-dimensional RAS method from the known structure of the annual tables and the quarterly totals.
Just as in the case of the regional matrices in Section \ref{sec:appSingle}, this approach does not ensure that the totals of the quarterly matrices are equal to the corresponding annual matrix.
In Table \ref{tab:comp}, we present the total difference between the annual tables and the sums of quarterly estimates by the two-dimensional RAS method.
The errors of the multidimensional RAS estimates totals are by definition always zero.

\begin{figure}
\centering
\includegraphics[width=14cm]{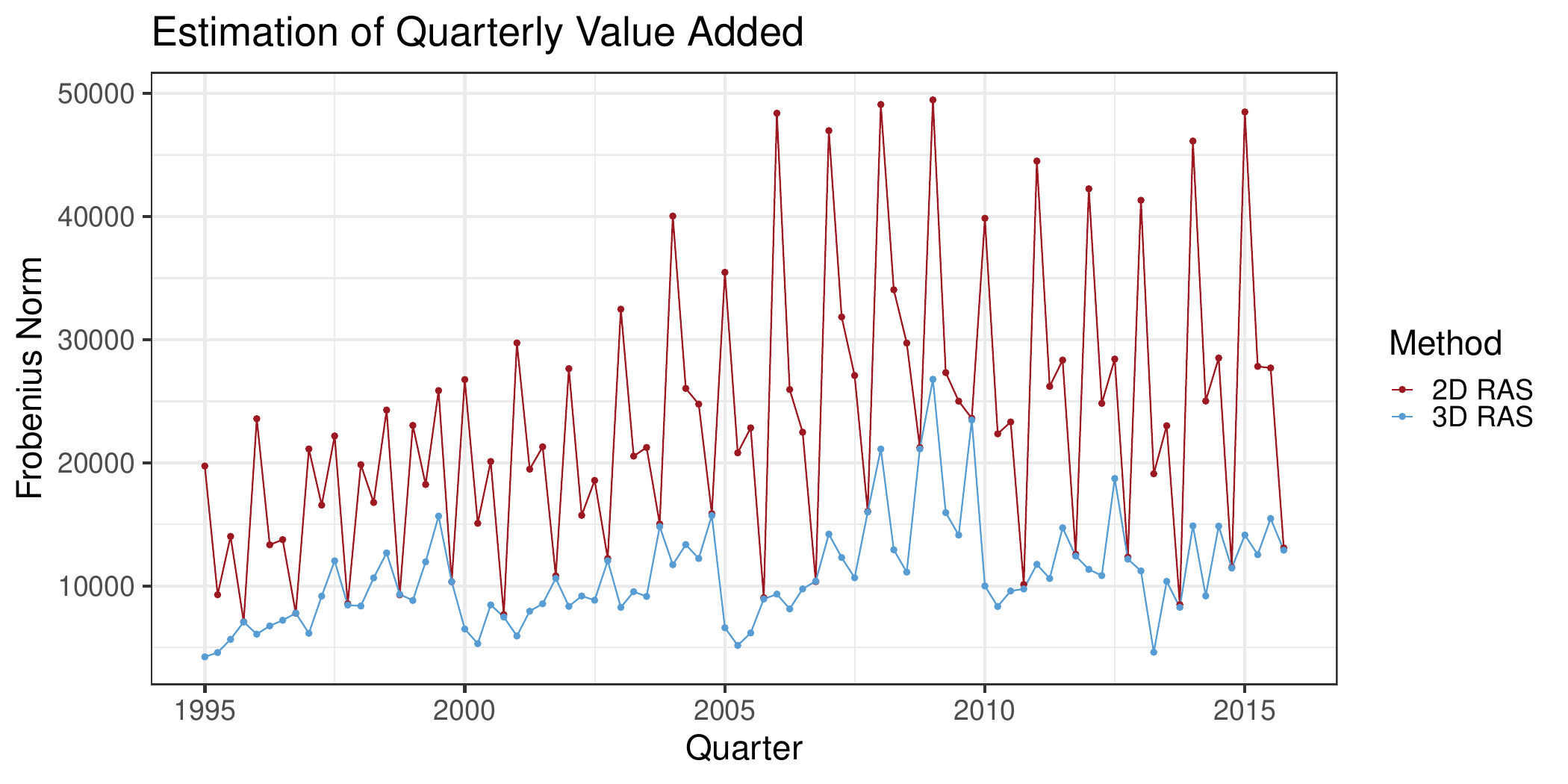} 
\caption{The Frobenius norm of the difference between the true quarterly value added of the Czech Republic from 1995 to 2015 and the two-dimensional and multidimensional estimates.}
\label{fig:quarterlyValue}
\end{figure}

We do not have the true quarterly input--output tables available.
However, we do know the quarterly value added (i.e., the last row of the estimated tables).
In the estimation, we treat the quarterly value added as unknown and consequently use its true value as a benchmark.
In Table \ref{tab:comp} and Figure \ref{fig:quarterlyValue}, we compare the estimates of the quarterly value added with its true values.
The multidimensional RAS method again outperforms the two-dimensional RAS method in terms of the Frobenius norm.
We can also observe that the errors are lowest for the years 1995, 2000, 2005, 2010, and 2013 when we know the annual input--output table.
In the other years, the errors are increasing in time, as the annual tables estimated from the last known annual table are less accurate.

The two-dimensional RAS method gives relatively accurate estimates for the fourth quarter of each year.
Moreover, the multidimensional RAS method gives almost the same estimate as the two-dimensional RAS method in the case of the fourth quarters.
We attribute this seasonality to the data compilation itself.
Namely, the acquisition of the quarterly national accounts data and their disaggregation into the level of detail that is in our tables.
First, the values of the annual national accounts are constructed as a sum of quarterly national accounts.
Next, when the annual surveys are completed, their values are split into quarters by benchmarking the original values of the quarterly national accounts.
This annualization then causes there to be almost a match in the fourth quarters.
Nevertheless, for the majority of years and quarters, the multidimensional RAS method gives far more accurate estimates.

\subsection{Domestic/Imported Input--Output Tables}
\label{sec:appDomimp}

The final application of the multidimensional RAS we present is the construction of the symmetric input--output tables themselves.
The third dimension represents the division into the domestic product use and the imported product use.
The input--output tables are typically designed by the Czech Statistical Office as follows.
First, the total table (TOT) including both the domestic use and the imported use is constructed.
Then the table for the domestic use (DOM) and the table for the imported use (IMP) are constructed from the TOT table using the two-dimensional RAS.
Here, a problem arises.
Although this approach ensures that the sum of the DOM and IMP tables have row and column sums equal to the TOT table, the individual elements of the sum of the DOM and IMP tables can be different from the elements of the TOT table.
This inconsistency causes the old national TOT table to be discarded and the new total table is created as the sum of DOM and IMP tables.
The purpose of this procedure for the construction is an expert revision of the input--output tables at the level of DOM and IMP tables.
For example, the DOM table is adjusted due to some minor changes in classification.

The aim of this experiment is to illustrate the situation in which only the total table and the column and row sums of the national and imported tables are known.
We estimate the DOM and IMP tables by each method and measure the differences.
The Frobenius norms of the total differences between TOT tables and the sums of DOM and IMP estimates over the years 1995, 2000, 2005, and 2010 are presented in Table \ref{tab:comp}.
We consider the multidimensional RAS method to be suitable for dividing the TOT table into the DOM and IMP tables, as the subsequent aggregation is consistent with the original TOT table.

\section{Conclusion}
\label{sec:con}

We have followed the work of \cite{Tilanus1976a} and \cite{Cole1992} on the multidimensional RAS method from both the theoretical and empirical perspectives.
Our paper has two main contributions:
\begin{itemize}
\item \textbf{Relation to the cross-entropy model.} In Theorem \ref{th:relation}, we prove that the solution of the multidimensional RAS method is the same as the solution of the multidimensional cross-entropy model.
We base our proof on the two-dimensional version of \cite{Lemelin2013}.
Although some of the related literature deals with the multidimensional extension of the RAS method, its theoretical properties have so far been mostly unverified.
\item \textbf{Empirical assessment.} In our extensive empirical study, we illustrate the benefits of the multidimensional RAS method.
Specifically, we have analyzed regional, quarterly, and domestic/imported industry-by-industry input--output tables of the Czech Republic.
Such a variety of applications allows us to assess the suitability of the multidimensional RAS method in both the spatial and temporal dimension.
The related literature, such as \cite{Temursho2021} and \cite{Valderas-Jaramillo2021}, focus mainly on spatial dimensions.
\end{itemize}

There are two key characteristics of the multidimensional RAS method which make it superior to the two-dimensional RAS method when disaggregating input--output tables:
\begin{itemize}
\item \textbf{Consistency.} The multidimensional RAS method ensures the consistency of the detailed tables with the overall input--output structure, in contrast with the two-dimensional RAS method.
The sum of the regional tables estimated by the multidimensional RAS method is always equal to those of the national table, the sum of the quarterly tables is equal to the annual table, and the sum of the domestic and imported tables is equal to the total table.
\item \textbf{Accuracy.} The multidimensional RAS method gives more accurate estimates than the two-dimensional RAS method when information on the totals is available in all dimensions.
In our empirical study, the tables estimated by the multidimensional RAS method are much closer to the true tables, in terms of the Frobenius norm.
The estimation of the Leontief inverse and quarterly value added also show a considerable increase in accuracy.
Only the application to the Isard's inter-regional input–output model shows comparable performance of both methods.
\end{itemize}

For these two reasons, we recommend not to omit higher dimensions of a given problem and to use the multidimensional RAS method when disaggregating tables to regional, quarterly, domestic/imported, and even more detailed input--output tables.
Alternatively, the multidimensional GRAS method of \cite{Temursho2021} and \cite{Valderas-Jaramillo2021} allowing for negative elements or the KRAS method of \cite{Lenzen2009} allowing for conflicting information as well as negative elements may be used.
Finally, note that these are fully expected findings in line with a general rule of \cite{DeMesnard2006} stating that ``introduction of accurate exogenous information into RAS improves the resulting estimates, and counterexamples should probably not be taken too seriously.''

\section*{Acknowledgements}
\label{sec:acknow}

We would like to thank Jaroslav Sixta, Michal Černý, and Alena Holá for their comments.

\section*{Funding}
\label{sec:fund}

The work of Vladimír Holý was supported by the Internal Grant Agency of the Prague University of Economics and Business under Grant F4/63/2016. The work of Karel Šafr was supported by the Czech Science Foundation under Grant 19-02773S.

%\bibliography{library.bib,data.bib}
%\bibliographystyle{mynatstyle}

\end{document}